\documentclass[a4paper]{article}
\pdfoutput=1

\usepackage[english]{babel}
\usepackage[utf8x]{inputenc}
\usepackage[T1]{fontenc}

\usepackage[a4paper,top=3cm,bottom=2cm,left=3cm,right=3cm,marginparwidth=1.75cm]{geometry}

\usepackage{amsmath}
\usepackage{graphicx}
\usepackage{lscape}
\usepackage{authblk}
\usepackage[colorinlistoftodos]{todonotes}
\usepackage[colorlinks=true, allcolors=blue]{hyperref}

\title{Game of Orbits: A Gaming Approach to Neptune's Discovery}

\author[1]{Siddharth Bhatnagar\thanks{siddharth.bhatnagar16ug@apu.edu.in}}
\author[2]{Jayant Murthy\thanks{jmurthy@yahoo.com}}
\affil[1]{Dept. of Physics, Azim Premji University}
\affil[2]{Indian Institute of Astrophysics}

\begin{document}
\maketitle

\begin{abstract}
\textit{One reason that the planet Neptune will be remembered for, is to do with the fact that it was the first planet whose existence was postulated before being observed. This was based on analysing deviations in Uranus' orbit. The present study gives a brief account of the history behind the discovery of Neptune, looking at the same, through the conflicting lenses of France and Britain.  With this context established, the study then investigates deviations in Uranus' orbit (experimentally found to range from 10,100 km to 21,276,000 km), under a host of different conditions and orientations of the perturbing body (Neptune/Jupiter). This was accomplished using Universe Sandbox 2, which is a physics-based simulation software, presented on the gaming platform, Steam. The paper then examines what these deviations correspond to, as viewed from Earth, finding that the deviation angle ranges between 0.04 arc-seconds and 3.22 arc-seconds. The paper in this way, demonstrates how using games like Universe Sandbox 2, can aid in approaching problems that would be considered difficult to visualize using a purely mathematical approach.}
\end{abstract}

\section{Introduction}

The discovery of the planet Neptune was a huge milestone in the field of astronomy as it was the first planet to be mathematically predicted before being observed. These predictions were essentially based on observing the anomalous motion of Uranus and the extent to which it was deviating from its theoretically defined orbit. The successful discovery of a planet based on gravitational perturbations caused by it on another planet further validated the Universal Law of Gravitation, as put forth by Sir Isaac Newton. The history of the discovery of the planet is, itself, very interesting, as it deals with the labours of brilliant mathematicians in an environment filled with doubt and uncertainty about the existence of an eighth planet. There are two sides to this story – a French one and a British one, each of whom would claim ownership over the discovery. It should not be forgotten here that the labours of both the key mathematicians involved in the discovery was of an extraordinary nature. They were both faced with an inverse problem, solvable by electronic computers today, but Herculean in the nineteenth century. 

This paper gives a brief account of the history of Neptune’s discovery, from both the British and the French perspectives. It then examines deviations in Uranus’ orbit due to Neptune under a host of different orientations and conditions, using Universe Sandbox 2. The paper then looks at what this deviation would look like from Earth. Lastly, it explores future research questions that could be answered using the likes of a game/simulation software such as Universe Sandbox 2.

\section{History}

By the late eighteenth century, seven planets were known to mankind, with Uranus having been discovered just two decades before the dawn of the new century. The discovery of this new planet in 1781 by Sir William Herschel, at once triggered the time and attention of astronomers. Uranus was, in fact, observable by the naked eye and was possibly first cataloged by the Greek astronomer, Hipparchus, in 128 BC and certainly had been observed at least twenty times before, between 1690 and 1771. Its faintness and its slow apparent motion in the sky meant that it had been mistaken to be a fixed star. 

Observations of the planet continued and, in 1790, the Academy of Sciences of Paris considered the nine years of accumulated observations sufficient, both in quality and quantity, for a decent approximation of the orbit of the planet and announced a prize for doing so. Jean Baptiste Joseph Delambre, a French mathematician and astronomer, duly calculated the elements of the planet’s orbit and with this knowledge, was able to construct tables for its motion, thereby carrying off the prize. For the next few years, all seemed fine, with the planet conforming to these tables. However, the planet soon began to assert its uniqueness and started deviating from the predicted positions in Delambre’s tables. At the time, this did not come as much of a surprise as it was believed that due to comparatively few observations and uncertain data, deviations from the table were bound to occur at some point in time. 

The next few years brought in more observations of Uranus and a few different hypotheses as to why Uranus could be deviating from the predictions. Enter Alexis Bouvard, a French astronomer, who in the 1820s, attempted to construct a new set of tables. He, like other astronomers of the time, was thoroughly troubled by the fact that Uranus was deviating from its theoretical orbit. What Bouvard found strangest was that though both, the earlier set of observations of the planet and those of more recent date could be perfectly represented by points on an ellipse, the two ellipses traced out were not the same. This was quite a peculiar problem! This problem led him to reluctantly conclude that the older observations were less precise and unreliable and he accordingly based his tables only on the new data. History repeated: for a few years, Uranus faithfully traced the orbit in accordance with Bouvard’s tables. But as before, this faithfulness soon turned to betrayal as Uranus began following a completely different curve, which consequently made Bouvard’s tables as unreliable as those of Delambre’s. One more attempt was made to construct new tables, but to no avail – the result was as unreliable as the tables of the predecessors. Seeing the failures of not one, not two, but three tables, the scientific community had an epiphany. In modern terminology, this could be thought of as an “It’s not me, it’s you” problem. The trouble lay not with the tables, but with the planet itself! \cite{Rines}

\subsection{The French Connection}

\begin{figure}[h]
\centering
\includegraphics[width=0.3\textwidth]{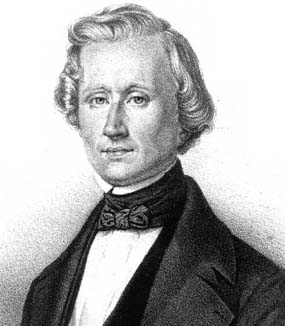}
\caption{\label{fig:Le_Verrier}Urbain Jean Joseph LeVerrier \cite{LeVerrier Portrait}}
\end{figure}

Enter Urbain Jean Joseph LeVerrier. This young mathematician had already found his place among the foremost of contemporary mathematical astronomers due to the high order of his analytical and practical mathematical skills. At the suggestion of his friend, Francois Arago, who was the then astronomical chief of France, the thirty-one year old LeVerrier dropped his research on comets and dedicated his time and effort thenceforth to solve what was one of the most puzzling astronomical problems of the nineteenth century. 

Firstly, LeVerrier thoroughly investigated Bouvard’s tables, to check for its accuracy. These investigations only went to prove that there were no computational errors in the tables. LeVerrier now became convinced that an unknown force was acting upon Uranus. As mentioned earlier, there were a few hypotheses as to why Uranus could be deviating from its theoretical orbit. One was that Newton’s Law of Gravitation was not accurate to distances as large as those between the Sun and Uranus. Another was that the scientific community had not taken into account the retarding influence of the ‘ether’ diffused through space (this was something that was disproved later on in the early twentieth century by the Michelson-Morley Experiment). Another suggestion was that Uranus had been dragged out of its orbit through collision or by the gravitational attraction of a passing comet. Yet another was that it was being perturbed by a large undiscovered satellite. Of all the different hypotheses, LeVerrier found the perturbation theory – that of a large undiscovered planet perturbing Uranus, as the most convincing.

He realized that this was an inverse problem (a problem that starts with the results and then arrives at the causes), or more technically, an inverse problem of perturbations. Though the problem could be approximately isolated to a three-body problem, involving the sun, Uranus and the perturbing body, the mathematical problem of locating the precise position of the object and its mass was still complex. A numerical approach was thus necessary in solving the problem. Now, since the distance of this unknown planet from the Sun was not known, it became clear that an intelligent guess of the same had to be made. The first approximation for this was made by using an empirical formula known as Titius-Bode’s Law, which held for the 5 planets of the ancient order. Using this, the unknown planet should have been twice as far from the Sun as the previous planet before it. After a grueling year, which included three months of LeVerrier not proceeding one bit in his researches out of pure discouragement, his calculations finally paid off and he could account for the perturbations in Uranus's orbit \cite{Rines}. On three occasions, LeVerrier presented his findings before the French Academy of Sciences. On November 10, 1845, he presented his preliminary investigations. On June 1, 1846, he presented before them the results of his first approximation. Lastly, on August 31 of the same year, he put forth his final results, mentioning the mass and orbital elements of the unknown planet and urged astronomers to devout some time in searching for it. To further motivate them, he remarked that the object would be of the eighth or ninth magnitude and could be resolved into a small disk, 3” in diameter. However, doubt triumphed over enthusiasm, resulting in inaction amongst many French astronomers. \cite{Smart}

Unable to excite any interest in the astronomers of his own country, LeVerrier then sent his results out by post to Johann Gottfried Galle, a German astronomer, at the Berlin Observatory. Galle received the post on 23 September and, unlike the French astronomers, was thrilled by the prospect of discovering a new planet in the skies. At the suggestion of a student, Heinrich Louis d’Arrest, Galle obtained a star map of the region of the sky in which the planet was to be, as per LeVerrier’s calculations. That very night, on the 23rd of September, 1846, Galle sat at the telescope and began calling out the configurations of the stars, as d’Arrest checked them off in the map. Around midnight, Galle found a star of the eighth magnitude in the telescope’s field of view, close to the theoretical position of LeVerrier’s planet. He called out its configuration to d’Arrest, who could not find it marked on the map. Voila! LeVerrier’s elusive planet had finally been found! \cite{Rines} After two more nights, when its position and movement were verified, Galle replied to LeVerrier’s letter, saying: “the planet whose place you have (computed) really exists!” \cite{Neptune Chronology} It was later on found by LeVerrier himself that his predicted position of the perturbing body was accurate to about 52’, i.e. less than a degree. This particular topic regarding LeVerrier's accuracy in predicting the position of the new planet has however been the subject of much debate and can be the focus of another paper. \cite{LeVerrier Error} It was also found that like its predecessor in the solar system, this planet had also been observed a few times before, twice by Galileo Galilei in 1612 and 1613, once by Jerome Lalande in 1795 and once by John Herschel in 1830, but all reported it as a star. \cite{O'Connor}

The scientific community was amazed and enthused by the discovery of a new planet "at the point of a pen". LeVerrier became a hero overnight. Astronomers hurriedly turned their telescopes to look at the elusive new planet, which had first been named LeVerrier and then changed to the neutral name, Neptune. While all this was happening merrily on the continent, unbeknownst to its scientific community, tension was brewing between France and Britain. \cite{Rines}

\subsection{The English Touch}

\begin{figure}[h]
\centering
\includegraphics[width=0.3\textwidth]{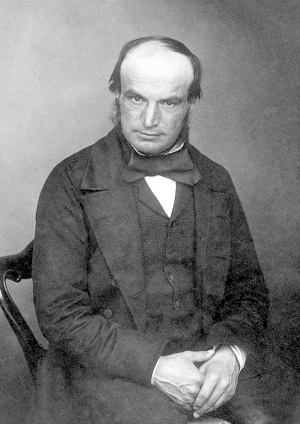}
\caption{\label{fig:johncouchadams}John Couch Adams \cite{Adams Picture}}
\end{figure}

Rewind back a few years from the discovery of Neptune and we find ourselves in St. Johns College, Cambridge. Enter John Couch Adams. During his second year of undergraduate college, he came to know of the mysterious motion of Uranus and was instantly drawn to it. He graduated in 1843 with the time and means to pursue the problem and began theorizing on the anomalous motion of the planet. \cite{Rines}

Adams declared that the orbital elements as described by Bouvard were wrong and that they needed to first be corrected. To these, the mass and the elements of the hypothetical planet would be added. To make it a simpler problem, a value was assumed for the mean distance between Uranus and the Sun. The same was done for the radius of the unknown planet’s orbit, which was to be in accordance with Bode’s Law. Both Adams and LeVerrier agreed that only a trial and error method would be appropriate to determine the radius of the planet’s orbit. \cite{Smart} As for the data, he decided to rely only on modern observations. His preliminary attempts were not very accurate, but nonetheless suggested to him that he should make slight improvements in his method. He soon received better data from the Astronomer Royal, Sir George Biddle Airy. This, coupled with an improved method, brought him to a much better result than before. Almost nine months prior to LeVerrier’s June 1846 address to the French Academy of Sciences, Adams, in September 1845 presented to James Challis, a professor at Cambridge, his paper with the elements of the unknown planet. As would happen with LeVerrier those many months later, this was not taken with much enthusiasm, neither by Challis nor by Airy, as it was considered “so novel a thing.” As one can imagine, this was very unfortunate for Adams, as it would give a headstart for LeVerrier and pave the way for him to claim total ownership over the discovery.

It was only in June of 1846, when LeVerrier published his paper containing the results of the first approximation, that Airy realized his folly. He found that LeVerrier had attained results very similar to those of Adams. Challis immediately left all tasks not urgent and began mapping the skies to search for the planet. Unfortunately, they did not have the map that Galle would go on to use, nor did they know of its existence – something which would have eased their labours greatly. On September 29, 1846, Challis received a copy of LeVerrier’s August paper and found that he too, just like Adams had commented that the object would be of the eighth or ninth magnitude. Without further ado, that very night he turned the telescope towards the planet’s theoretical position in the sky. There, one "star" out of some three hundred present, seemed to possess a disk. This was Adam’s planet! Unfortunately for Adams, history would know it as LeVerrier’s planet - Neptune, thanks to Airy and Challis’ inaction of nine months.

In the next many months, tensions grew between Britain and France with regard to who should be given the rights to the discovery. It was certain that Adams had in reality, theoretically discovered Neptune before LeVerrier, but the fact of the matter was that LeVerrier had established his ownership over the discovery because of formal publication. Foreign astronomers, from a more neutral stance of the situation, agreed with this. They also realised that in the case of Adams, the fault lay with Challis and Airy. Adams did not take any part in the whole controversy, nor did he blame anyone. \cite{Rines} To shield Airy and Challis from immense criticism, he even went on to comment: “I could not expect however that practical astronomers, who were already fully occupied with important labours, would feel as much confidence in the results of my investigations, as I myself did.” \cite{British Steal}

\begin{figure}[h]
\centering
\includegraphics[width=10 cm]{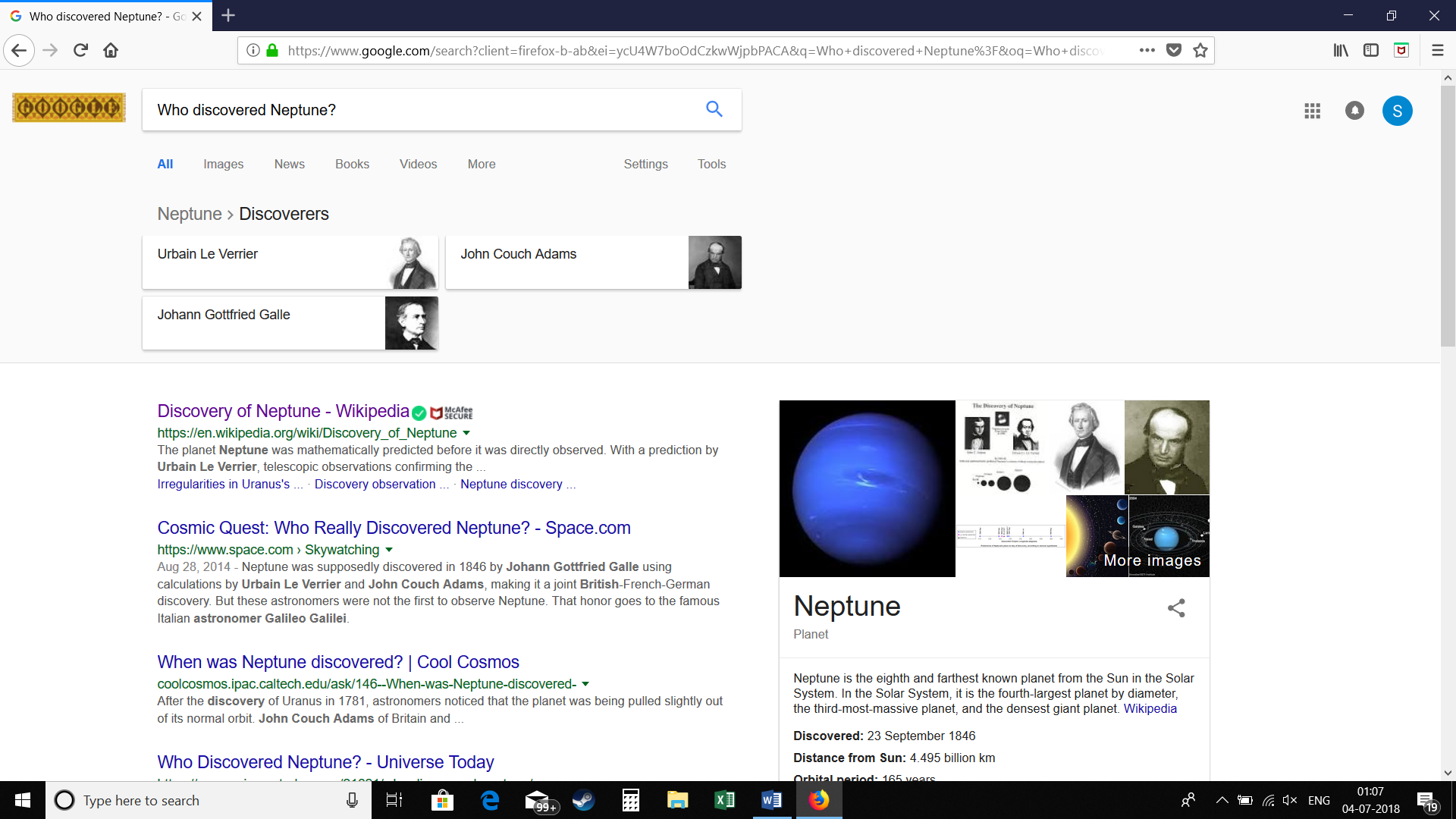}
\caption{\label{fig:Screenshot_1_.png}Google Search for 'Who discovered Neptune?' \cite{Google Search}}
\end{figure}

Fast-forwarding 172 years to 2018, a search on Google with the words, “Who discovered Neptune?” yields an answer befitting to the labours of the astronomers involved in the discovery (shown in Figure 3). The answer being - Urbain Le Verrier, John Couch Adams and Johann Gottfried Galle.

\section{Calculating Neptune's orbit}

We do not have the mathematical skill necessary to recreate the calculations of LeVerrier and Adams and have turned to Universe Sandbox 2, which according to their website "is a physics-based space simulator. It merges gravity, climate, collision and material interactions \ldots” \cite{Universe Sandbox 2} We used only the gravity part of this simulator to explore the effects of adding a planet near the orbit of Uranus.

The following takes us through the steps involved in measuring deviations in the osculating orbit of Uranus, using Universe Sandbox 2. The osculating orbit refers to the orbit that a body around a central object would have had, if it was not being perturbed in any manner. This section will then look at how the angle subtended by the deviation with respect to Earth was calculated, i.e. the angle that the deviation makes, as viewed from Earth.

\subsection{Method}

As mentioned earlier, the exact calculations carried out by LeVerrier and Adams is beyond the scope of this paper as the issue presented is a complex inverse problem, which would have multiple solutions and would need a knowledge of perturbative theory. In this limited project, we modeled only three body systems (Sun, Uranus, Neptune/Jupiter) and neglected orbital eccentricities and interactions of the other planets. We are aware of the fact that both Jupiter and Saturn would, in reality, impact the orbit of Uranus but limited ourselves to the three body problem in this work. We will not delve into the details of the program here but, rather, briefly touch upon the methods used.

The first step was to start a new simulation with the Sun in the center of the simulation’s Cartesian system and Uranus at a distance of 19.2 AU ($2.9 \times 10^{6}$ km) on the X axis, corresponding to the real mean distance of Uranus from the Sun. The orbit was defined to the XZ plane. It should be noted here that this served as the starting condition for all the successive cases and sub-cases and we will call this the \textbf{base simulation.} The simulation was played at a reasonable speed of 11 months/sec and the distance of Uranus from the Sun was noted down at 5 different instances: Position 1a, corresponding to the starting position on the positive X axis, Position 2 on the positive Z axis, Position 3 on the negative X axis, Position 4 on the negative Z axis and Position 1b, corresponding to the ending position on the positive X axis. Tracing each of these positions of Uranus in the order mentioned above, as a function of time, would essentially correspond to completing one entire orbit (though unclosed) around the sun. 

\begin{figure}[h]
\centering
\includegraphics[width=10 cm]{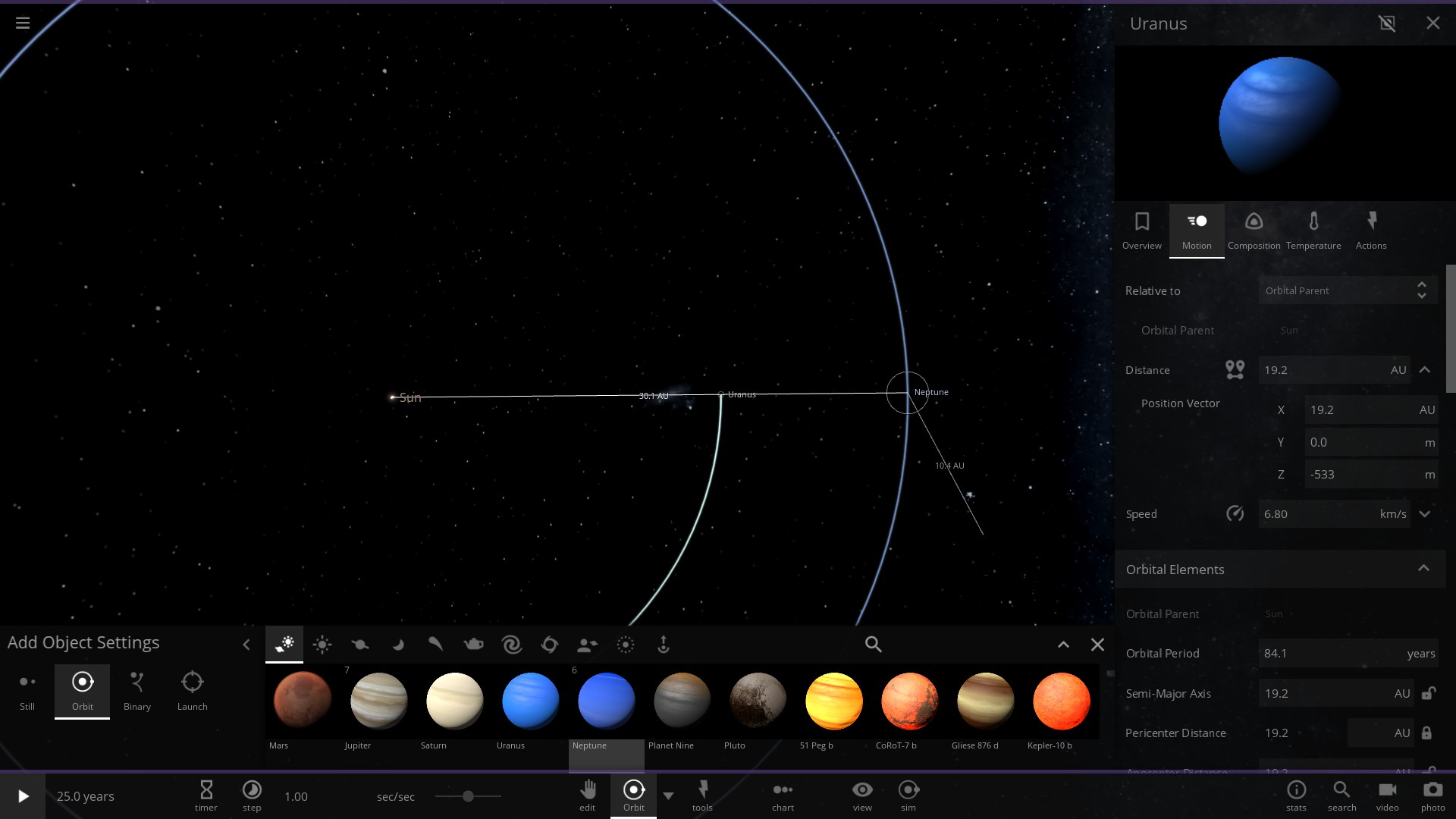}
\caption{\label{fig:20180704005156_1.jpg}Adding Neptune at 30.1 AU in the simulation at 0$^{\circ}$ to Sun - Uranus system}
\end{figure}

\begin{figure}[h]
\centering
\includegraphics[width=10 cm]{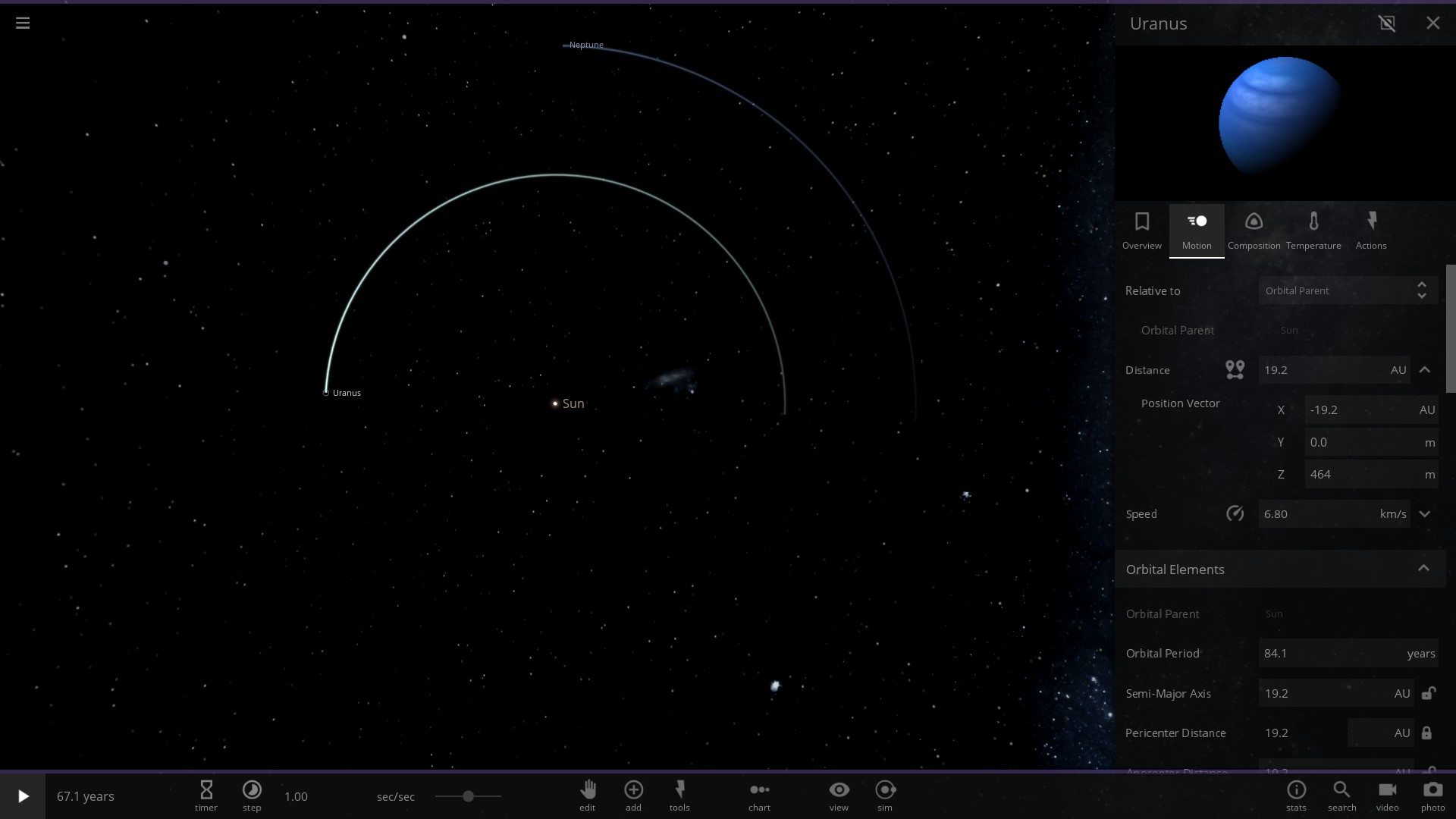}
\caption{\label{fig:20180704005523_1.jpg}Pausing simulation as Uranus reaches the negative Z axis (notice 1:2 resonance in orbits of Uranus and Neptune)}
\end{figure}

Three different systems were worked upon, each having four different initial orientations (0º, 90º, 180º, 270º) of the perturbing body (Neptune/Jupiter) with respect to the Sun – Uranus system. Eg: For 90º, the perturbing body started at the distance as described by the system itself, but at a position 90º ahead of the + X axis, i.e. on the + Z axis. This was mainly done to account for the fact that in reality, the perturbing body could be at any arbitrary position with respect to the conditions of the base simulation. These systems and sub-cases will be explored in more detail in a case by case scenario.

\begin{itemize}
\item \textbf{Case 1 (Sun – Neptune = 30.1 AU)}
\end{itemize}
This case involved adding Neptune at a distance of 30.1 AU from the Sun (shown in Figure 4), which corresponds to its real mean distance from the Sun over one period of revolution. This was added to the paused base simulation, thereafter which the simulation was played. For each sub-case (orientation), the five different positions (1a, 2, 3, 4, 1b) were noted down.\footnote{Note: The same procedure was carried out for the bottom two cases as well.}

\begin{itemize}
\item \textbf{Case 2 (Sun – Neptune = 38.4 AU; in accordance with Bode’s Law)}
\end{itemize}
As discussed earlier, the astronomers of the nineteenth century estimated that the orbit of Neptune would extend to a distance double that of Uranus’, in accordance with Bode’s Law. \cite{Rines}\cite{Smart} This would put Neptune at a distance of 38.4 AU. This is what was established as the starting conditions of this case.

\begin{itemize}
\item \textbf{Case 3 (Sun – Jupiter = 30.1 AU)}
\end{itemize}
For this case, Jupiter was placed instead of Neptune at a distance of 30.1 AU. The starting conditions for this case were identical to that of Case 1, except for the fact that instead of Neptune, Jupiter was placed. This was done primarily to observe just how much Jupiter would have influenced Uranus’ orbit, had it been present in Neptune’s place.

Table 1 contains the distances of Uranus from the Sun at each of the pre-defined positions, for each of the cases and sub-cases. Along with this, the deviation in distance for each of the positions is also given. For position 1a, 2, 3 and 4, this is calculated by calculating the difference between the distance at that position and the theoretical distance for that position. For position 1b, the deviation is calculated by calculating the difference between the distance at that position with the distance at the starting condition (2,871,975,200 km). \textbf{Positive deviation} values indicate that the deviation is outward, away from the Sun, while \textbf{negative deviation} values indicate that the deviation is inward, towards the Sun. The same has been indicated in Figure 6.

\begin{figure}[h]
\centering
\includegraphics[width=15 cm]{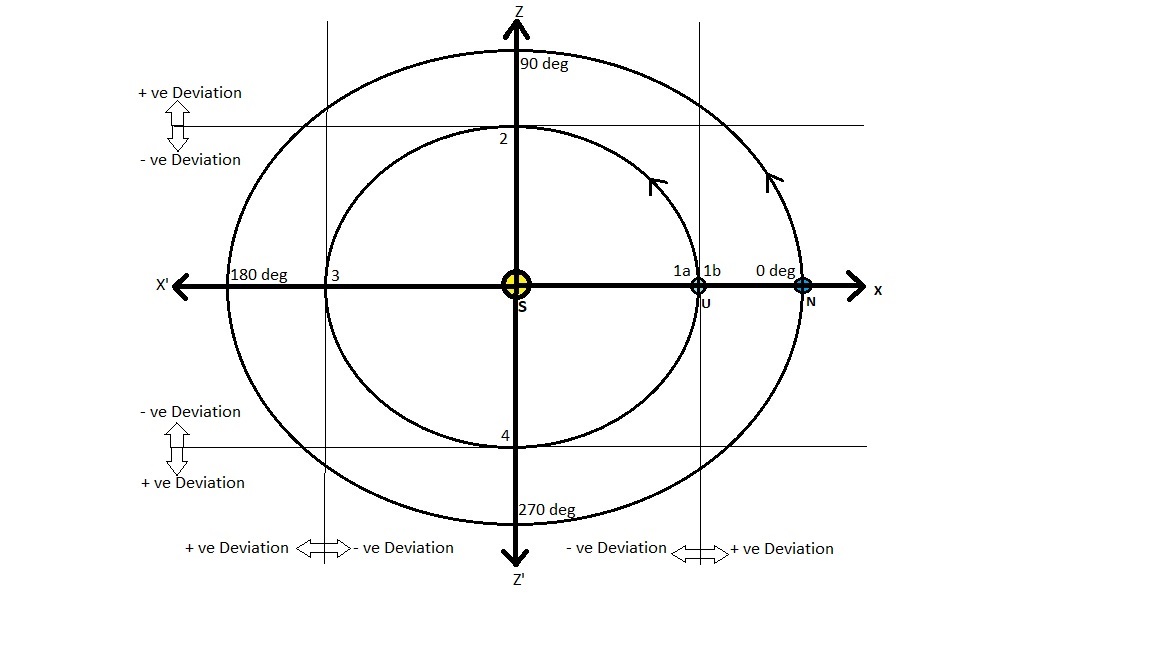}
\caption{\label{fig:Figure_1_BaseDiagram.jpg}Visualising deviations in the orbit of Uranus}
\end{figure}

\subsection{Deviations as viewed from Earth:}
For this, the orbital periods of Earth and Uranus were first calculated, using the orbital period formula, based on Kepler’s Third Law:
$$T=2\pi\sqrt{\frac{a^3}{GM}}$$
Where, 

T = time period of orbiting body

a = semi-major axis distance of orbiting body

G = Universal Gravitational Constant

M = Mass of Sun 
\begin{figure}[h]
\centering
\includegraphics[width=12 cm]{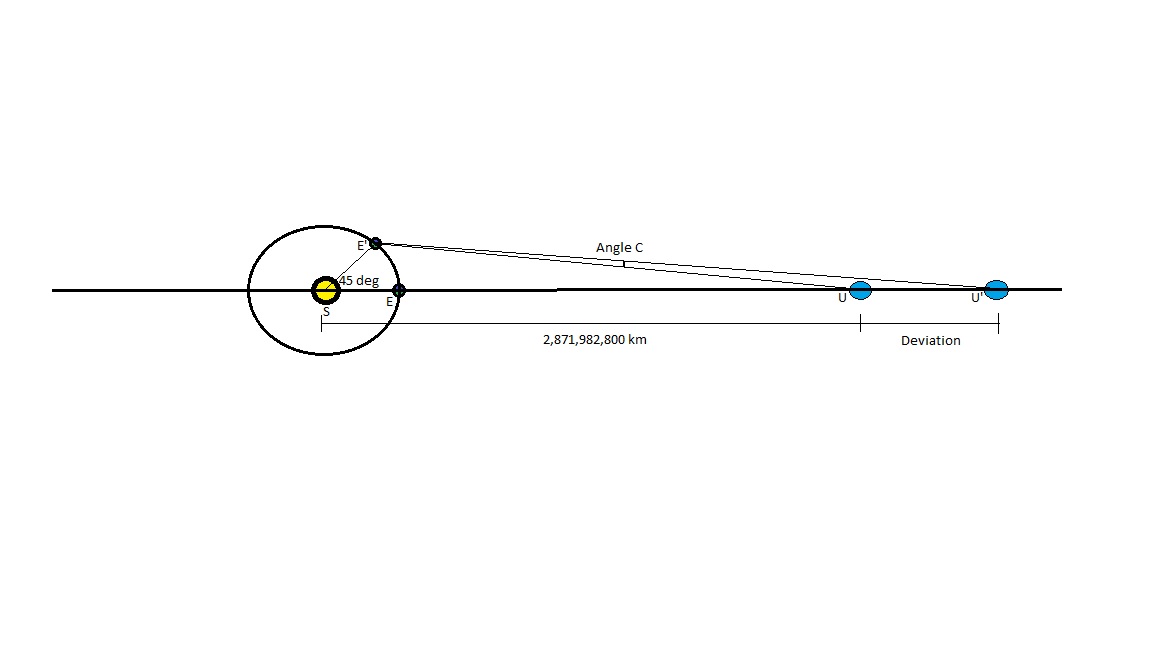}
\caption{\label{fig:Figure_2_Deviation_Angle.jpg}Deviation and the corresponding angle as viewed from Earth}
\end{figure}

The ratio of Uranus’ time period to that of Earth’s yielded a value of 84.125. \cite{Planets} This meant that Uranus took 84.125 Earth years to orbit the Sun once. This result was very convenient in a sense as it implied that at the end of 84.125 years, while Uranus would have just come back to the + X axis, Earth would have started its 85th orbit of the sun, of which it would have completed only 0.125 of an entire revolution. This corresponds to an angle of 45º with the Sun and the + X axis, and consequently, Uranus. Now, with respect to Uranus, it should be understood that in this case, the two items of interest are the theoretical position that Uranus would be at after completing one revolution (Position 1b of the base simulation) and the actual position that Uranus is at after completing one revolution (given by Position 1b of the respective sub-case). The difference in these values correspond to the deviation at Position 1b. Earth’s position 45º into its orbit, Uranus’ actual position and Uranus’ theoretical position together form a triangle, as shown in Figure 7. The angle subtended at Earth by this triangle corresponds to the deviation in the planet’s orbit, as viewed from Earth.

Table 2 has in it imported Position 1a and Position 1b values from Table 1. In addition to this, the deviation for Point 1b as described above is mentioned. As we can see from Figure 7, side E’U forms one of the sides of the triangle E’SU. Here, E’U was calculated using the Cosine Rule as the other parameters of the triangle were known. This includes side E’S, which is the distance between the Sun and Earth, after Earth having completed 84.125 revolutions, side SU which is the theoretical distance that Uranus should be at after having completed one revolution and angle E’SU, which is 45º. As one can see from the figure, the distance E’U stays constant. The formula used here, as per the Cosine Rule, is as follows:
$$E'U=\sqrt{E'S^2+SU^2-2(E'S)(SU)\cos\theta}$$
Where, $\theta = 45$º

In a similar manner, using the cosine rule, side E’U’ and consequently, the angle subtended at Earth due to the deviation distance UU’, can be calculated from triangle E’UU’. To calculate the angle, we solve for angle C, which is given by:
$$C =\arccos{[\frac{E'U^2+E'U'^2-UU'^2}{2(E'U)(E'U')}]}$$

Table 2 contains the different deviation angle values for each of the sub-cases. It should be noted here that the deviation values and angles for the first item (theoretical – yellow) do not really make sense as those do not involve true deviations due to any perturbing body. This will be discussed in the Results and Discussion section.

\begin{landscape}

\section{Results and Discussion}

\begin{figure}[h]
\centering
\includegraphics[width=20 cm]{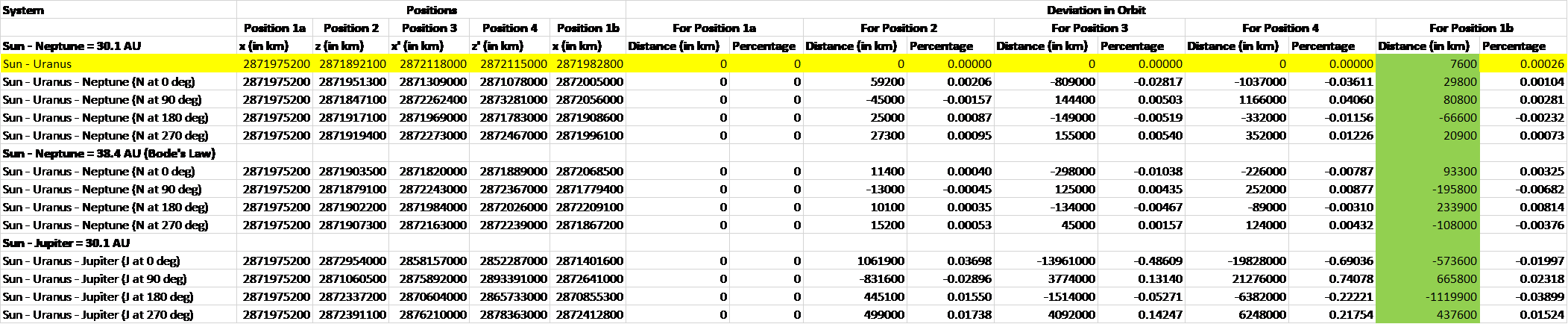}
\caption{\label{fig:Table_1_Deviations.png}Table 1 - Deviations}
\end{figure}

\begin{figure}[h]
\centering
\includegraphics[width=11.2 cm]{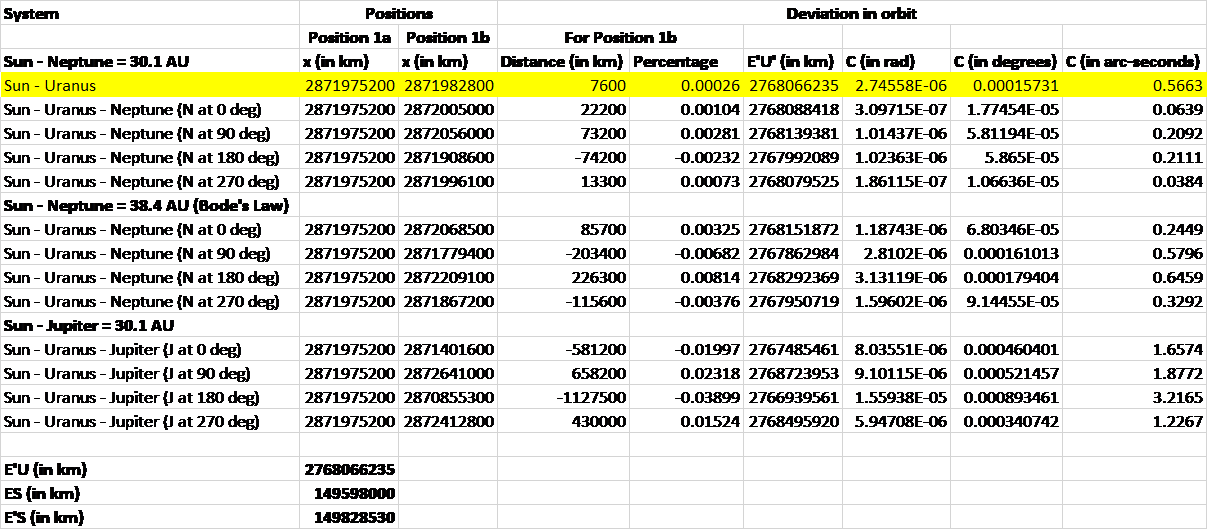}
\caption{\label{fig:Table_2_Deviation_Angles.png}Table 2 - Deviation Angles as viewed from Earth}
\end{figure}

\end{landscape}

One of the first things apparent from the deviations in Table 1 is that the deviations for Position 1a are consistently zero. This is purely because of the fact that the starting position for each case was precisely the same point on the axis (2,871,975,200 km) for all trials. It can also be seen that for Systems 1 and 3, the deviations for each position increase significantly across the positions, with the deviation at Position 2 being the least and that at Position 4 being the most. It should be noted here that between the orbits of Uranus and Neptune, there exists approximately 1:2 resonance. This implies that for every one revolution of Uranus, Neptune has covered only half a revolution around the Sun. For the simple case of analysing just one orbit of Uranus around the Sun, this implies that by the time Uranus reaches Position 2 from Position 1a, and assuming Neptune also started from 0 deg (on the + X axis), Neptune would be somewhere in the middle of the first quadrant, but as it is outward of Uranus’ orbit, it would pull Uranus towards it, thereby resulting in positive deviation. Negative deviation occurs consistently at Position 2 for when Neptune starts at 90 deg (on the + Z axis) because of the fact that by the time Uranus reaches Position 2, Neptune would be somewhere in the middle of the second quadrant, pulling Uranus towards it, i.e. inward of Uranus’ orbit itself. These are crude explanations for justifying the same and a much more thorough explanation can be given by analysing the vector components of each of the bodies. What can however be said with much greater ease here is to appreciate the difference in deviation when Jupiter is placed in the system instead of Neptune. The deviations here are now consistently of the order of one or two factors greater than with Neptune. This is primarily because of the fact that Jupiter is massive (almost 20 times more massive than Neptune itself! \cite{Planets}) Another aspect to appreciate here is that between Systems 1 and 3, though the deviations are factors more, the signs of the deviation remain the same for each position and orientation, going on to show the independence of mass on the direction of pull on the perturbed body. An analysis of the like is more complex for System 2, as it does not show 1:2 resonance like the former two systems and would require us to analyse the problem with the use of vectors.

When we look at Table 2, one thing that becomes apparent is that the deviation values for this table are not the same as those for Table 1. This is because these deviation values are the differences between the actual position of Uranus after an entire revolution and the theoretical position of Uranus after the same. The deviation values in Table 1 are essentially the difference in the starting and ending positions of Neptune. It should be noted here that the values of E’S and that of SU are different from ES and the actual distance between SU. There could be two reasons for this, one being the fact that the orbits as defined in the simulation undergo precession, just like in the real-world. Another could be just out of pure simulation error. One of the noticeable aspects of this table is the magnitude of increase in the deviations between Systems 1 and 3 and consequently, the increase in the deviation angle C, between Systems 1 and 3. This deviation angle C is the angle of deviation as viewed from Earth. For System 1, the deviation angle varies between 0.06" to 0.21". While these are in fact small deviations, when the simulation is run over many more cycles, these would amount to significant differences in the deviation angle and as a consequence, would result in a difference in its position in the sky as viewed from a telescope on Earth. The deviation angle for Jupiter is of a much more appreciable magnitude, lying between 1.23” and 3.22”. For just one simulation cycle, such a deviation would register as a significant difference in position in the sky, as viewed from Earth.

\section{Conclusion}

As we can see, with the use of what many would even call a game, Universe Sandbox 2 serves a good purpose in estimating deviations of such small magnitude, when used in an appropriate manner. It is however true that the simulation is not free of errors. Significant difficult-to-quantify errors can come up during the running of the simulation itself, especially when simulation time speeds are increased, as parameters defined by the program become less precise. In addition to this, not accounting for the effects of the other planets, especially the Jupiter-Saturn system, is a big approximation to make, but this was done to make the simulation run more smooth and error-free. All the more, this allows us to observe the effects of an isolated Uranus – Neptune system. It would however be very interesting to get to measuring deviations in Uranus’ osculating orbit, by placing all the solar system planets in the simulation. With this, it would be great to see what measure of difference in deviation exists between the two simulations (with Neptune and with all other solar system planets). In addition to this, it would be interesting to note what the deviation, as measured from the software, would look like from Earth, in the form of celestial coordinates. This would involve the use of rotation matrices and transformations between celestial and Cartesian coordinates. Using this as a baseline, future work could be done to try and map out these deviations, as they would have appeared in the night sky of 23 September 1846. With this done, it would be fascinating to try and draw out a comparison between this and what Galle/Challis observed in September of 1846, based on the help provided by LeVerrier/Adams.

A fascinating topic of contemporary research and something that is closely in relation with the theme of this paper is to do with the search to find the much talked about Planet Nine through indirect detection. This is being carried out using perturbation theory and with the help of today's electronic computers. Estimates on the extent of its orbit range from between 300-400 AU \cite{Lower Estimate} all the way to 600-700 AU \cite{Anton}\cite{Upper Estimate 1}\cite{Upper Estimate 2}\cite{Upper Estimate 3}. It is interesting to note here that the latter estimates happen to coincidentally concur with estimates as put forth by Bode's Law (614.4 AU)! If Planet Nine really is found within this range, it could go on to strengthen this law and consequently make it a decent empirical tool to find and estimate the semi-major axes of planets revolving around other stars. \cite{Anton}

Lastly, it is worth to mention here that using games like Universe Sandbox 2, which is essentially a physics-based simulator software, really does go a long way in approaching and viewing problems that might be difficult to visualize otherwise. These problems could range from measuring deviations in orbits, to measuring how much force would be required to deflect an asteroid falling onto Earth, to measuring changes in the spin of a body before and after a collision, to measuring temperature variations by extending orbits. The prospects truly are plentiful. An approach such as this aids in visually being able to see events occur, and at the same time, for the more scientifically inclined, work with supplementary data provided by the software to check if theory meets fact.

\section{Acknowledgements}

SB takes this opportunity to thank Professor Proteep C. Mallik, for getting him in touch with Dr. Murthy and for his comments during the course of the research. He is also grateful to Dr. Suri Venkatachalam for helping him contact IIA. His sincerest thanks goes out to Dr. Jayanth Vyasanakere and Dr. Anand Shrivastava in helping him to think about the problem at hand in certain novel ways. Last, but not the least, he would like to thank Dr. Rajaram Nityananda for taking the time to read through some of the content with regard to the paper and offer his expert comments on the same.

\end{document}